# Polar Coding for Parallel Gaussian Channel

David Tse[+], Bin Li[*], Kai Chen[*]
[+]Dept. of Electrical Engineering, Stanford University, USA, dntse@stanford.edu
[*]Dept. Communication Technology Research, Huawei Technologies, Shenzhen, P. R. China
binli@huawei.com, kaichen@ieee.org

*Abstract*—In this paper, we propose a Polar coding scheme for parallel Gaussian channels. The encoder knows the sum rate of the parallel channels but does not know the rate of any channel. By using the nesting property of Polar code, we design a coding/decoding scheme to achieve the sum rates.

*Keywords-Polar codes; Paralle Gaussian channels.*

## I. INTRODUCTION

Polar codes are a major breakthrough in coding theory [1]. They can achieve Shannon capacity with a simple encoder and a simple successive cancellation decoder when the code block size is large enough. For a Gaussian channel $W$ whose capacity is $R$, if the transmitter knows that the channel capacity is $R$, we can design a capacity-achieving polar code of (long) block length $N$ designed for this channel to reach the rate $R$. For the parallel Gaussian channel consisting of $M$ independent Gaussian channels with capacity of $R_m$, where $1 \le m \le M$, respectively, if the transmitter knows all rates: $R_1, R_2, ..., R_M$, we can design $M$ capacity-achieving Polar codes with rates $R_1, R_2, ..., R_M$ to reach the sum rate capacity $R_1 + R_2 + ... + R_M$ of this parallel channel. One open problem is that if the transmitter only knows the sum rate but does not have any information about the rate of any one of the parallel channels, can we still achieve this sum rate capacity? Without loss of generosity, we assume that the rates of $M$ parallel channels are $r_1, r_2, ..., r_M$ respectively and the sum rate is one, i.e., $r_1 + r_2 + ... + r_M = 1$. As shown in Fig. 1, we need to design a coding scheme which encodes $N$ bits and send them over $M$ parallel channels. The receiver can jointly and successfully decode $N$ bits.

In section II, we review the encoding of Polar codes and its nesting property and in section III we propose a new Polar coding scheme for parallel channel. Finally we draw some conclusions

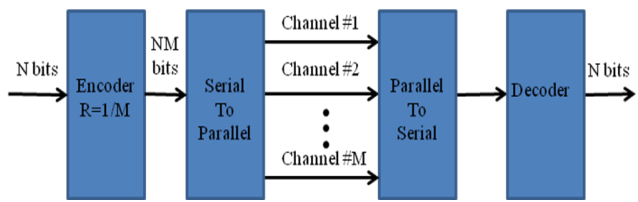

Figure 1.  Joint encoding/decoding for *M* parallel channels.

## II. POLAR CODES AND ITS NESTING PROPERTY

### A. Encoding of Polar Codes

Let $F = \begin{bmatrix} 1 & 0 \\ 1 & 1 \end{bmatrix}$, $F^{\otimes n}$ is a $N \times N$ matrix, where $N = 2^n$, $\otimes n$ denotes $n$th Kronecker power, and $F^{\otimes n} = F \otimes F^{\otimes(n-1)}$. Let the *n*-bit binary representation of integer $i$ be $b_{n-1}, b_{n-2}, ..., b_0$. The *n*-bit representation $b_0, b_1, ..., b_n$ is a bit-reversal order of $i$. The generator matrix of polar code is defined as $G_N = B_N F^{\otimes n}$, where $B_N$ is a bit-reversal permutation matrix. The polar code is generated by

$$x_1^N = u_1^N G_N = u_1^N B_N F^{\otimes n} \qquad (1)$$

where $x_1^N = (x_1, x_2, ..., x_N)$ is the encoded bit sequence, and $u_1^N = (u_1, u_2, ..., u_N)$ is the encoding bit sequence. According to the principle of Polar design, these encoding bits $(u_1, u_2, ..., u_N)$ have different reliabilities, and these *N* bits are divided into two subsets according to their reliabilities. The top *K* most reliable bits are used to send information and the rest are frozen bits set to zeros. Let *S* represent the subset containing the information bit indexes, then the coding rate is $R = |S|/N$, where $|S|$ is the size of *S*.

### B. Nesting Property of Polar Codes

We use the nesting property of Polar codes [2] to construct rate-compatible Polar code as follows: we divide the *N* encoding bits into *Q* subsets: $S_1, S_2, …, S_K$, where $S_1$ contains the indices of the top *N/Q* most reliable bits, $S_2$ contains the indices of the second top *N/Q* most reliable bits, …, and $S_K$ contains the indices of the most unreliable *N/Q* bits. For a channel *W* whose capacity is $R = m/Q$, where $1 \le m \le Q$. We can design a capacity-achieving polar code of (long) block length *N* designed for a channel *W* by choosing the information bit indices as $\sum_{k=1}^{m} \cup S_k$.

## III. POLAR CODING FOR TWO PARALLEL CHANNELS

We illustrate our design using a simple example of 2 parallel channels as shown in Fig. 2. Let $Q=4$ and the encoding bits are divided into four subsets $S_1, S_2, S_3, S_4$ according to reliability. Every input $N$ bits are divided into four sub-blocks and are placed in $S_1, S_2, S_3, S_4$ positions, respectively. Note that they are encoded in four Polar codes. For example, the Polar coding for the first channel is as follows: the first $N$ input bits are divided into $a_1, a_2, a_3, a_4$. $a_1$ is placed in $S_1$ of the first Polar code, $a_2$ is placed in $S_2$ of the second Polar code, …, and $a_4$ is placed in $S_4$ of the fourth Polar code. The Polar coding for the second channel is the same as the first channel but uses the reversal order of $a_1, a_2, a_3, a_4$.

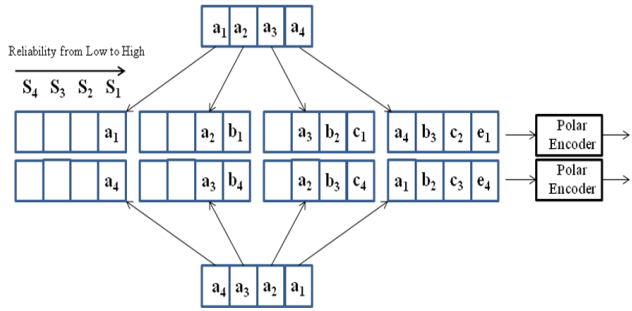

Fig. 2. Polar coding for 2 parallel channels.

When the two channel rates are $r_1 = m/4$ and $r_2 = 1 - m/4$, where $m$ is an integer and $0 \leq m \leq 4$, we can successfully decode all information bits. The reason is as follows: the first $m$ codes in the first channel can be successfully decoded due to their rates are less than or equal to the channel rate $r_1 = m/4$, we can obtain information bits $a_1, a_2,..., a_m$; The first 4-m codes in the second channel can be successfully decoded due to their rates are less than or equal to the channel rate $r_2 = 1 - m/4$, we can obtain information bits $a_4, a_3,..., a_{m+1}$; Therefore we can obtain all information $a_1, a_2,..., a_4$. After we cancel them from first 4 codes for both of the two channels, we can repeat the above process to decode the next $b_1, b_2,..., b_4$, and so on, until all bits are decoded. Note that the coding rate for the first Polar code is $R=1/4$, the second is $R=2/4$ and the third is $R=3/4$, the following codes have full rates of $R=1$. There is some overhead or rate loss for the first three Polar codes, but when the number of the transmitted codes is large enough, this overhead is negligible.

Fig.3 shows a general structure of Polar coding for two parallel channels, where the $N$ encoding bits are divided into $Q$ subsets according to reliability. When the two channel rates are $r_1 = m/Q$ and $r_2 = 1 - m/Q$, where $m$ is an integer and $0 \leq m \leq Q$, we can successfully decode all information bits as we explain above. The first $m$ codes in the first channel can be successfully decoded due to their rates are less than or equal to the channel rate $r_1 = m/Q$, we can obtain information bits $a_1, a_2,..., a_m$; the first Q-m codes in the second channel can be successfully decoded due to their rates are less than or equal to the channel rate $r_2 = 1 - m/Q$, we can obtain information bits $a_Q, a_{Q-1},..., a_{m+1}$; Therefore we can obtain all information $a_1, a_2,... a_Q$. After we cancel these decoded bits from all involved codes, we can repeat the above process to decode the next $b_1, b_2,...b_Q$, and so on, until all bits are decoded.

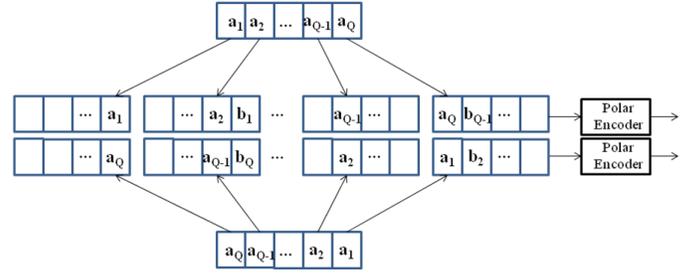

Fig. 3. A General Structure of Polar coding for 2 parallel channels.

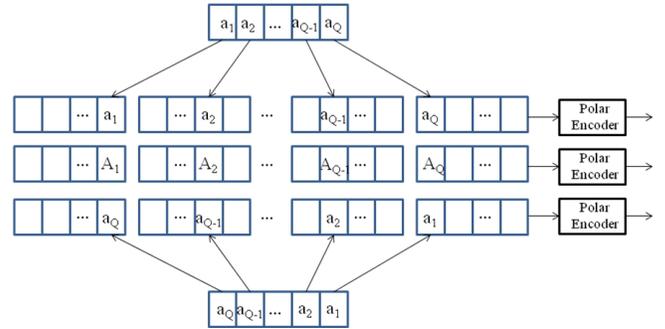

Fig. 4. A General Structure of Polar coding for 3 parallel channels.

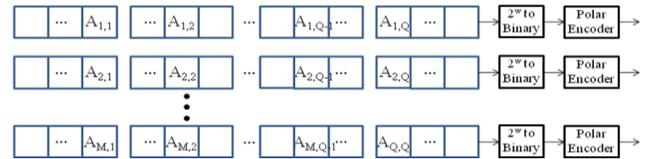

Fig. 5. A General Structure of Polar coding for M parallel channels.

## IV. POLAR CODING FOR $M$ PARALLEL CHANNELS

Fig.4 shows a general structure of Polar coding for three parallel channels, where $(A_1, A_2,..., A_Q) = (a_1, a_2,..., a_Q) F^{\otimes n}$, and both $A_k$ and $a_k$ are $N/Q$-bit sequences, $1 \leq k \leq Q$. When the three channel rates are $r_1 = k_1/Q$, $r_2 = k_2/Q$ and

$r_3 = k_3/Q$, where $k_1, k_2, k_3$ are integers $0 \leq k_1, k_2, k_3 \leq Q$, $k_1 + k_2 + k_3 = Q$, we can successfully decode all information bits. The reason is as follows: We can successfully decode $a_1, a_2, \ldots a_{k_1}$, $A_1, A_2, \ldots A_{k_2}$, and $a_Q, a_{Q-1}, \ldots a_{Q-k_3+1}$ from the first, the second and the third channel, respectively. Let $G$ be a sub-matrix of $F^{\otimes n}$, which contains the first $k_2$ columns and $k_2$ rows (from $(k_1+1)$-th to $(k_1+k_2)$-th columns) of $F^{\otimes n}$, then we have $(A_1, A_2, \ldots, A_{k_2}) = (a_{k_1+1}, a_{k_1+2}, \ldots a_{k_1+k_2}) \times G$. $G$ of size $k_2 \times k_2$ has full rank (see the Appendix I), therefore we can solve $(a_{k_1+1}, a_{k_1+2}, \ldots a_{k_1+k_2})$. Now all of $(a_1, a_2, \ldots a_Q)$ are obtained. After we cancel them from all involved codes, we repeat the above process to decode $b_1, b_2, \ldots, b_Q$ and so on.

When the number of parallel channels $M>3$, we will use UDM (Universal Decodable Matrix) [3] for our design. Unfortunately there are no binary UDMs for $M>3$. Let the input information $(a_1, a_2, \ldots, a_Q)$ be on $GF(2^w)$, $a_k$ is a $\frac{N}{wQ}$-symbol sequence, $1 \leq k \leq Q$. Let $H_k$ be UDM with size of $\frac{N}{wQ} \times \frac{N}{wQ}$, $1 \leq k \leq M$, and $(A_{k,1}, A_{k,2}, \ldots A_{k,Q}) = (a_1, a_2, \ldots a_Q) H_k$, where $A_{k,j}$ is a $\frac{N}{wQ}$-symbol sequence, where $1 \leq k \leq M$ and $1 \leq j \leq Q$, when the $M$ channel rates are $r_1 = k_1/Q$, $r_2 = k_2/Q$, ..., and $r_M = k_M/Q$, where $k_1, k_2, \ldots, k_M$ are integers $0 \leq k_1, k_2, \ldots, k_M \leq Q$, $k_1 + k_2 + \cdots + k_M = Q$, we can decode $(A_{1,1}, A_{1,2}, \ldots A_{1,Q})$ from the first channel, $(A_{2,1}, A_{2,2}, \ldots A_{2,Q})$ from the second channel,..., and $(A_{M,1}, A_{M,2}, \ldots A_{M,Q})$ from the $M$-th channel. According to the property of UDM, we can decode $(a_1, a_2, \ldots, a_Q)$.

## V. CONCLUSIONS

In this paper, we propose a Polar coding scheme for parallel Gaussian channel. The $M$ parallel channels have the capacity rates: $r_1, r_2, \ldots, r_M$ with the sum rate $r_1 + r_2 + \ldots + r_M = 1$. The transmitter has the only information that the sum rate is one and does not know any rate of the channels. In order to deal with the nonknown channels, every input $N$ bits are transformed by UDMs and encoded by $Q$ Polar codes for each channel. The decoder decodes every $Q$ Polar codes from each channel and obtains part of these $N$ bits and part of their transformed bits. Due to the property of UDM, we solve equations to obtain all $N$ bits.

## REFERENCES

[1] E. Arıkan, "Channel polarization: A method for constructing capacity achieving codes for symmetric binary-input memoryless channels," IEEE Trans. Inform. Theory, vol. 55, pp. 3051–3073, July 2009.
[2] B. Li, D. Tse, K. Chen and H. Shen, "Capacity-Achieving Rateless Polar Codes," available as online as arXiv: 1508.03112v1.
[3] A. Ganesan and P. O. Vontobel, "On the Existence of Universal Decodable Matrices," available as online as arXiv: 0601066v1

## APPENDIX I

Theorem: Sub-matrix $G$ of $F^{\otimes n}$ is full rank.

Proof: Let $c(i, j)$ be the $i$-th row and the $j$-th column element of $F^{\otimes n}$, where $1 \leq i, j \leq N$. The sub-matrix $G$ of $F^{\otimes n}$ is defined as follows: the $i$-th row and the $j$-th column element of $G$ is $c(i+i_0, j)$, where $1 \leq i, j \leq K$, $K$ is an positive integer and $K < N$, $0 \leq i_0 \leq N-K$.

1) When $n=3$, $N=8$, any sub-matrix $G$ is full rank;
2) Induction hypothesis that any sub-matrix $G$ of $F^{\otimes n}$ is full rank. We need to prove that any sub-matrix $G$ of $F^{\otimes(n+1)}$ is full rank, where $F^{\otimes(n+1)}$ has a size of $(2N) \times (2N)$, but we only need to consider the following four cases: a) $N \leq K < 2N$; b) $N/2 \leq K \leq N$, $0 \leq i_0 \leq N/4$; c) $N/2 \leq K \leq N$, $N/4 < i_0 \leq N/2$; d) $N/2 \leq K \leq N$, $N/2 \leq i_0 \leq 3N/4$; Other cases can be easily proved from the induction hypothesis.

*Case I.* $N \leq K < 2N$

The sub-matrix $G$ (ABCD) is shown in Fig. 6, which contains two identical matrices IJHD and KLCG, and three zero matrices: MFQN, FBLK and OKGP. After matrix "XOR" operation, FBCG=FBCG $\oplus$ AEHD, $G$ contains zero matrix KLGC as shown in Fig. 7.

Since the square matrix NQKO is full rank (this can be easily proved), we use it to "cancel" matrix QRLK to make it a zero-matrix by column combinations of FBCG and MFGP as shown in Fig. 8.

The matrices AETS and FBRQ are identical, AETS can be "cancel" by XOR operation: AEHD=AEHD $\oplus$ FBCG. STIJ can be "canceled" by NQKO. Now we have three square matrices FBRQ, NQKO and IOPD as shown in Fig. 9. From the induction hypothesis, FBRQ and IOPD are full rank.

*Case II.* $N/2 \leq K \leq N$, $0 \leq i_0 \leq N/4$

The sub-matrix $G$ (ABCD) is shown in Fig. 10, which contains two identical matrices IJLK and GHNM, and two zero matrices OBHG and MNCP. After matrix XOR, AEFD=OBCP $\oplus$ AEFD, the new form $G$ is as shown in Fig. 11.

The AEJI and KLFD form a matrix QRST which is full rank according to induction hypothesis. We use it to cancel EOGJ and LMPF by column combination between AEFD and EOPF and obtain a new form of $G$ as shown in Fig.13, where the matrix JHNL is a column-shifted of $F^{\otimes(n-1)}$, which is full rank.

*Case III.* $N/2 \leq K \leq N$, $N/4 \leq i_0 \leq N/2$

The sub-matrix *G* (ABCD) is shown in Fig. 14, which contains a zero matrix JFHK. The matrices IBFJ and KHCL form a square matrix MNOP which is full rank according to induction hypothesis. The matrices AIJE and GKLD can be "canceled" by IBFJ and KHCL with linear combination of columns to obtain a new of *G* as shown in Fig. 15. The matrix EJHG is the same as $F^{\otimes(n-1)}$, which is full rank.

*Case IV.* $N/2 \leq K \leq N$, $N/2 \leq i_0 \leq 3N/4$

The sub-matrix *G* (ABCD) is shown in Fig. 16, which contains three zero matrices: IJNM, OGCL and HPKD. We use MNOP to "cancel" NFGO and obtain a new form of *G* as shown in Fig. 17. The matrix JBFN is full rank according to the induction hypothesis, we use it to cancel matrix AIME and use MNOP to cancel EMPH, and obtain a new form of *G* containing JBNF, IJNM and HPDK as shown in Fig. 18, where JBNF and HPDK full rank according to induction hypothesis.

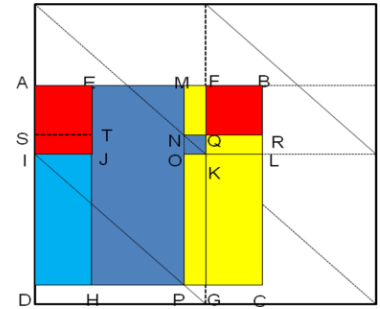

Fig. 8. New form of Sub-matrix *G* in case I.

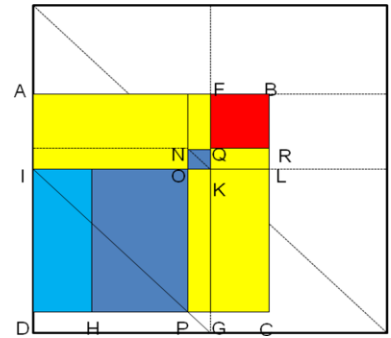

Fig. 9. New form of Sub-matrix *G* in case I.

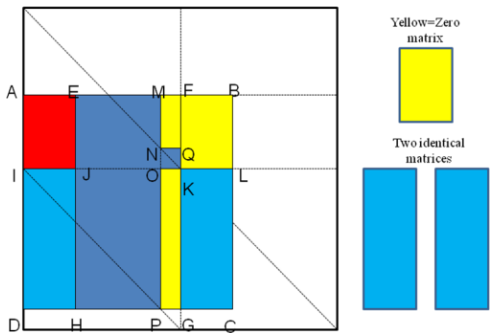

Fig. 6. Sub-matrix *G*: ABCD in $F^{\otimes(n+1)}$ in case I.

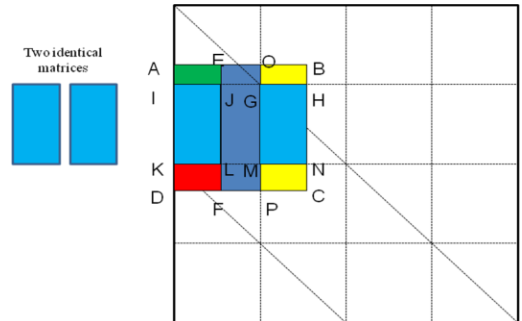

Fig. 10. Sub-matrix *G*: ABCD in $F^{\otimes(n+1)}$ in case II.

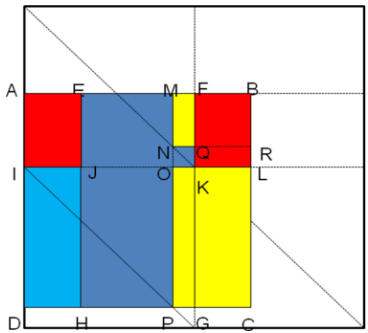

Fig. 7. New form of Sub-matrix *G* in case I.

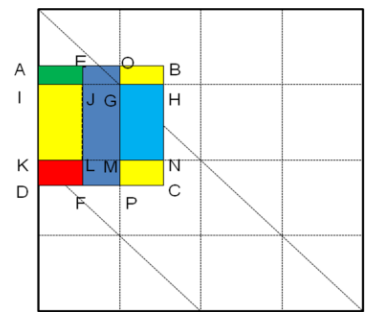

Fig. 11. New form of Sub-matrix *G* in case II.

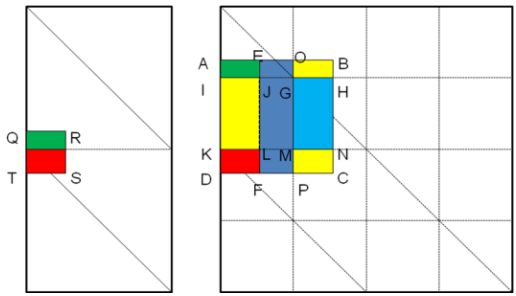

Fig. 12. New form of Sub-matrix *G* in case II.

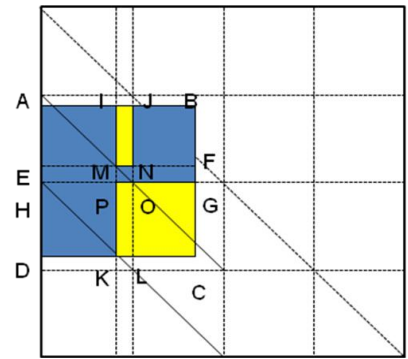

Fig. 16. Sub-matrix *G*: ABCD in $F^{\otimes(n+1)}$ in case IV.

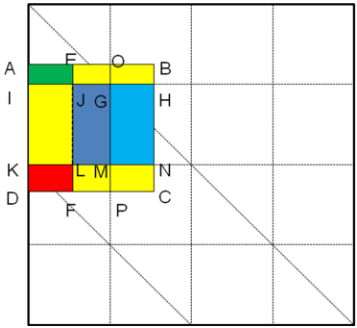

Fig. 13. New form of Sub-matrix *G* in case II.

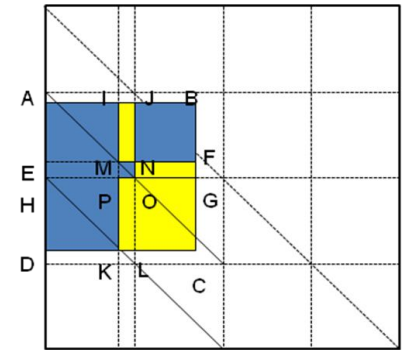

Fig. 17. New form of Sub-matrix *G* in case IV.

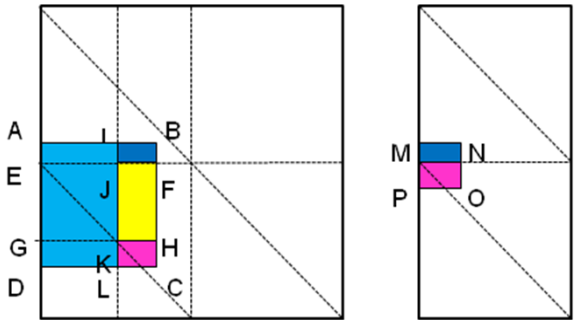

Fig. 14. Sub-matrix *G*: ABCD in $F^{\otimes(n+1)}$ in case III.

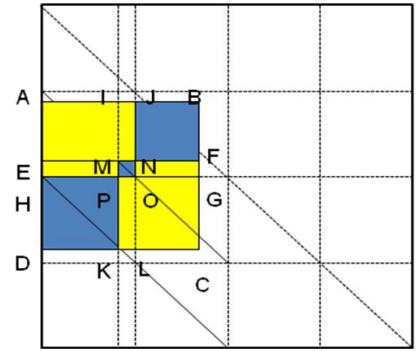

Fig. 18. New form of Sub-matrix *G* in case IV.

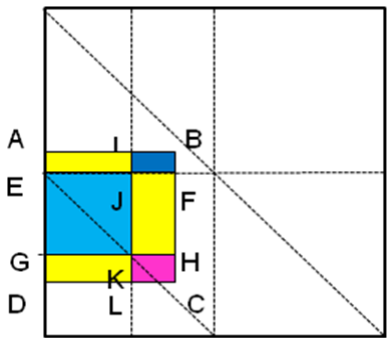

Fig. 15. New form of Sub-matrix *G* in case III.